\documentclass[journal=acsccc,manuscript=article]{achemso}
\makeatletter\if@twocolumn\PassOptionsToPackage{switch}{lineno}\else\fi\makeatother

\usepackage{tabulary,amsfonts,amssymb,amsbsy,amsmath}
\usepackage{ifluatex}
\ifluatex
\usepackage{fontspec}
\defaultfontfeatures{Ligatures=TeX}
\usepackage[]{unicode-math}
\unimathsetup{math-style=TeX}
\makeatletter
\renewcommand{\acs@author@fnsymbol}[1]{{\expandafter\acs@author@fnsymbol@aux\expandafter{\number#1 }}}
\makeatother
\else 
\usepackage[T1]{fontenc}
\usepackage[utf8]{inputenc}
\fi 
\usepackage[version=3]{mhchem}
\let\savephone\phone 
\SectionNumbersOn
\usepackage{url,multirow,morefloats,floatflt,cancel,tfrupee}
\makeatletter

\AtBeginDocument{\@ifpackageloaded{textcomp}{}{\usepackage{textcomp}}}
\makeatother
\usepackage{colortbl}
\usepackage{xcolor}
\usepackage{pifont}
\usepackage[nointegrals]{wasysym}
\makeatletter

\def\mcWidth#1{\csname TY@F#1\endcsname+\tabcolsep}

\def\cAlignHack{\rightskip\@flushglue\leftskip\@flushglue\parindent\z@\parfillskip\z@skip}
\def\rAlignHack{\rightskip\z@skip\leftskip\@flushglue \parindent\z@\parfillskip\z@skip}

\usepackage{ifxetex}
\ifxetex\else\if@twocolumn\usepackage{dblfloatfix}\fi\fi

\AtBeginDocument{
\expandafter\ifx\csname eqalign\endcsname\relax
\def\eqalign#1{\null\vcenter{\def\\{\cr}\openup\jot\m@th
  \ialign{\strut$\displaystyle{##}$\hfil&$\displaystyle{{}##}$\hfil
      \crcr#1\crcr}}\,}
\fi
}

\AtBeginDocument{%
  \@ifpackageloaded{endfloat}%
   {\renewcommand\efloat@iwrite[1]{\immediate\expandafter\protected@write\csname efloat@post#1\endcsname{}}}{\newif\ifefloat@tables}%
}%

\def\BreakURLText#1{\@tfor\brk@tempa:=#1\do{\brk@tempa\hskip0pt}}
\let\lt=<
\let\gt=>
\def\processVert{\ifmmode|\else\textbar\fi}

\@ifundefined{subparagraph}{
\def\subparagraph{\@startsection{paragraph}{5}{2\parindent}{0ex plus 0.1ex minus 0.1ex}%
{0ex}{\normalfont\small\itshape}}%
}{}

\newcommand\role[1]{\unskip}
\newcommand\aucollab[1]{\unskip}
  
\@ifundefined{tsGraphicsScaleX}{\gdef\tsGraphicsScaleX{1}}{}
\@ifundefined{tsGraphicsScaleY}{\gdef\tsGraphicsScaleY{.9}}{}
\def\checkGraphicsWidth{\ifdim\Gin@nat@width>\linewidth
	\tsGraphicsScaleX\linewidth\else\Gin@nat@width\fi}

\def\checkGraphicsHeight{\ifdim\Gin@nat@height>.9\textheight
	\tsGraphicsScaleY\textheight\else\Gin@nat@height\fi}

\def\fixFloatSize#1{}%\@ifundefined{processdelayedfloats}{\setbox0=\hbox{\includegraphics{#1}}\ifnum\wd0<\columnwidth\relax\renewenvironment{figure*}{\begin{figure}}{\end{figure}}\fi}{}}
\let\ts@includegraphics\includegraphics

\def\inlinegraphic[#1]#2{{\edef\@tempa{#1}\edef\baseline@shift{\ifx\@tempa\@empty0\else#1\fi}\edef\tempZ{\the\numexpr(\numexpr(\baseline@shift*\f@size/100))}\protect\raisebox{\tempZ pt}{\ts@includegraphics{#2}}}}

\AtBeginDocument{\def\includegraphics{\@ifnextchar[{\ts@includegraphics}{\ts@includegraphics[width=\checkGraphicsWidth,height=\checkGraphicsHeight,keepaspectratio]}}}

\DeclareMathAlphabet{\mathpzc}{OT1}{pzc}{m}{it}

\def\URL#1#2{\@ifundefined{href}{#2}{\href{#1}{#2}}}

\def\UrlOrds{\do\*\do\-\do\~\do\'\do\"\do\-}%
\g@addto@macro{\UrlBreaks}{\UrlOrds}

\edef\fntEncoding{\f@encoding}

\makeatother

\newif\ifmultipleabstract\multipleabstractfalse%
\let\phone\savephone

\title{COMBIgor: data analysis package for combinatorial materials science}

\author{Kevin R. Talley}

\affiliation[National Renewable Energy Laboratory]{National Renewable Energy Laboratory\unskip, 15013 Denver West Pkwy\unskip, Golden\unskip, CO\unskip, 80401\unskip, USA}
\alsoaffiliation[Colorado School of Mines]{Department of Metallurgical and Materials Engineering \unskip, Colorado School of Mines\unskip, 1500 Illinois St.\unskip, Golden\unskip, CO\unskip, 80401\unskip, USA}

\author{Sage R. Bauers}
\affiliation[National Renewable Energy Laboratory]{National Renewable Energy Laboratory\unskip, 15013 Denver West Pkwy\unskip, Golden\unskip, CO\unskip, 80401\unskip, USA}

\author{Celeste L. Melamed}
\affiliation[National Renewable Energy Laboratory]{National Renewable Energy Laboratory\unskip, 15013 Denver West Pkwy\unskip, Golden\unskip, CO\unskip, 80401\unskip, USA}
\alsoaffiliation[Colorado School of Mines]{Department of Physics\unskip, Colorado School of Mines\unskip, 1500 Illinois Street\unskip, Golden\unskip, CO\unskip, 80401\unskip, USA}

\author{Meagan C. Papac}
\affiliation[National Renewable Energy Laboratory]{National Renewable Energy Laboratory\unskip, 15013 Denver West Pkwy\unskip, Golden\unskip, CO\unskip, 80401\unskip, USA}
\alsoaffiliation[Colorado School of Mines]{Department of Metallurgical and Materials Engineering \unskip, Colorado School of Mines\unskip, 1500 Illinois St.\unskip, Golden\unskip, CO\unskip, 80401\unskip, USA}

\author{Karen N. Heinselman}
\affiliation[National Renewable Energy Laboratory]{National Renewable Energy Laboratory\unskip, 15013 Denver West Pkwy\unskip, Golden\unskip, CO\unskip, 80401\unskip, USA}

\author{Imran Khan}
\affiliation[National Renewable Energy Laboratory]{National Renewable Energy Laboratory\unskip, 15013 Denver West Pkwy\unskip, Golden\unskip, CO\unskip, 80401\unskip, USA}

\author{Dennice M. Roberts}
\affiliation[National Renewable Energy Laboratory]{National Renewable Energy Laboratory\unskip, 15013 Denver West Pkwy\unskip, Golden\unskip, CO\unskip, 80401\unskip, USA}
\alsoaffiliation[University of Colorado Boulder]{Department of Mechanical Engineering\unskip, University of Colorado Boulder\unskip, Boulder\unskip, CO\unskip, 80309\unskip, USA}

\author{Valerie Jacobson}
\affiliation[National Renewable Energy Laboratory]{National Renewable Energy Laboratory\unskip, 15013 Denver West Pkwy\unskip, Golden\unskip, CO\unskip, 80401\unskip, USA}
\alsoaffiliation[Colorado School of Mines]{Department of Metallurgical and Materials Engineering \unskip, Colorado School of Mines\unskip, 1500 Illinois St.\unskip, Golden\unskip, CO\unskip, 80401\unskip, USA}

\author{Allison Mis}
\affiliation[National Renewable Energy Laboratory]{National Renewable Energy Laboratory\unskip, 15013 Denver West Pkwy\unskip, Golden\unskip, CO\unskip, 80401\unskip, USA}
\alsoaffiliation[Colorado School of Mines]{Department of Metallurgical and Materials Engineering \unskip, Colorado School of Mines\unskip, 1500 Illinois St.\unskip, Golden\unskip, CO\unskip, 80401\unskip, USA}

\author{Geoff L. Brennecka}

\affiliation[Colorado School of Mines]{Department of Metallurgical and Materials Engineering \unskip, Colorado School of Mines\unskip, 1500 Illinois St.\unskip, Golden\unskip, CO\unskip, 80401\unskip, USA}

\author{John D. Perkins}

\affiliation[National Renewable Energy Laboratory]{National renewable Energy Laboratory\unskip, 15013 Denver West Pkwy\unskip, Golden\unskip, CO\unskip, 80401\unskip, USA}

\author{Andriy Zakutayev}

\affiliation[National Renewable Energy Laboratory]{National Renewable Energy Laboratory\unskip, 15013 Denver West Pkwy\unskip, Golden\unskip, CO\unskip, 80401\unskip, USA}

\email{Andriy.Zakutayev@NREL.gov}

\usepackage{float}

\begin{document}

\begin{abstract}
Combinatorial experiments involve synthesis of sample libraries with lateral composition gradients requiring spatially-resolved characterization of structure and properties. 
Due to maturation of combinatorial methods and their successful application in many fields, the modern combinatorial laboratory produces diverse and complex data sets requiring advanced analysis and visualization techniques. 
In order to utilize these large data sets to  uncover new knowledge, the combinatorial scientist must engage in data science.  
For data science tasks, most laboratories adopt common-purpose data management and visualization software.
However, processing and cross-correlating data from various measurement tools is no small task for such generic programs. 
Here we describe COMBIgor, a purpose-built open-source software package written in the commercial Igor Pro environment, designed to offer a systematic approach to loading, storing, processing, and visualizing combinatorial data sets. 
It includes (1) methods for loading and storing data sets from combinatorial libraries, (2) routines for streamlined data processing, and (3) data analysis and visualization features to construct figures. 
Most importantly, COMBIgor is designed to be easily customized by a laboratory,
group, or individual in order to integrate additional instruments and data-processing algorithms. 
Utilizing the capabilities of COMBIgor can significantly reduce the burden of data management on the combinatorial scientist.
\end{abstract}
    
\section{Introduction}
High-throughput experimental (HTE) research, which consists of synthesis and characterization of specimens with spatially varied processing parameters, has been established as an effective strategy for exploring and optimizing materials.
These methods, also known as combinatorial science, have proven useful for discovery and optimization of materials for diverse applications, including catalytic materials\unskip~\cite{374021:8323623}, electronic and functional ionic materials\unskip~\cite{374021:8323688}, polymer coating materials\unskip~\cite{374021:8323845}, sensing materials\unskip~\cite{374021:8323773}, and biomaterials\unskip~\cite{374021:8323602}. 
With the success of combinatorial science, a diverse set of high-throughput synthesis and characterization instruments has been developed and demonstrated\unskip~\cite{374021:8323731},  facilitating rapid generation and analysis of HTE materials data. 

The ability to input and analyze data is typically the rate-limiting process in HTE methods\unskip~\cite{Maier2007}.
The HTE researcher must process, analyze, and correlate a large amount of data in order to understand the variations within each combinatorial library, then understand how trends correlate between libraries. 
This process is inefficient, as the experiments are put on hold so that scientists can wade through the data before continuing.
Programmatic task handling ensures that effort is not unnecessarily wasted on routine tasks, such as loading, sorting, labeling, and tracking data. 
Accessible data and flexible analysis algorithms are needed for processing and analyzing various types of data and effectively interpreting results.  

There have been several combinatorial software packages presented in literature. 
One example is CombiVIEW\unskip~\cite{374021:8337235}, a MatLab package for the analysis and visualization of x-ray diffraction patterns generated from combinatorial approaches. 
Another package was written to improve HTE data organization and sharing for sensor materials.\cite{Frantzen2005}
Developments like these have significant \textit{depth}. 
They handle one type of data or perform one specific task exceptionally well. 
The lacking element is \textit{breadth}, the ability to handle many different types of HTE data in many different ways, so that the data can be more efficiently stored, compared, shared, and presented. 
Thus, a package is needed that has sufficient breadth, while being flexible enough to enable the user to develop depth for their own processes and data types.

Here we present COMBIgor - a data analysis package for Igor Pro intended to aid in the management, processing, and cross-correlation of multi-dimensional data sets collected by mapping combinatorial libraries on spatially-resolved measurement instruments. 
The core functionality was intentionally designed to enable flexibility in data types and experimental setup, while enabling the user to expand usability by adding new measurement instruments and new data analysis algorithms. 
Below, we present an overview of the functionality and architecture of the open-source COMBIgor package. 
This package can reduce the burden of data management of combinatorial experiments, thus encouraging its adoption by combinatorial researchers, groups, and institutions.

\section{Overview}

\subsection{COMBIgor History}
Development of this software package started in the early 2000s by materials scientists at the National Renewable Energy Laboratory (NREL) in Golden, Colorado alongside their emerging thin-film HTE methodology. 
From 2000 to 2010, several software add-ons (referred to as "procedures" in Igor Pro) were developed in order to load, store, and display combinatorial data from individual instruments, such as a Bruker D8 Discover x-ray diffractometer (XRD) and MaXXi 5-pin x-ray fluorescence (XRF) spectrometer.
In 2010, this collection of procedures was extended to include custom optical and electronic property mapping instruments.\cite{Zakutayev2013b} 
Most importantly, these different procedures were integrated into a package allowing processing and cross-correlation of different types of data, as cited in two published manuscripts\unskip~\cite{zakutayev_perkins_parilla_widjonarko_sigdel_berry_ginley_2011,PhysRevB.85.085204}.
Since that time, the package has continued to grow as new and continuing users have added new instruments and new functionality, and has since been used to process data and generate figures for a large number of publications. Here, we intentionally cite the publications in which this package was mentioned, but not cited, so that future readers of these publications can find the present manuscript and better understand the software package used.
\cite{Han2018,Han2018a,Siol2018,Arca2018,Siol2018a,Han2017,Bikowski2017,Welch2017,Fioretti2017,Rajbhandari2017,Siol2016,Ndione2016,Garten2016,Bikowski2016,Siol2016a,Fioretti2016,Caskey2016,Welch2016,Fioretti2015,Baranowski2015,Welch2015,Zakutayev2015,Peng2015,Subramaniyan2015,Caskey2015,Welch2015a,Mercado2014,Baranowski2014,Zakutayev2014,Subramaniyan2014,Zakutayev2013,Zawadzki2013,Zakutayev2013a,Peng2013,Zakutayev2013b,Cloet2012,Paudel2011}

Around 2016, the authors started considering improved methods of data management, usability, and visualization, as well as the potential for the package to be extended by outside users and developers. 
In the two years that followed, the collection of Igor procedures was restructured and optimized into an interactive and flexible Igor package for loading, storing, and processing combinatorial data, and was given the name \textit{COMBIgor} (COM-bee-gore). 
This newer version has already been used for data analysis and preparation of figures in several recent manuscripts.\cite{ 374021:8586428,374021:8586454,Bauers2018,sage2,CM_PRM,DMR_latest} 
Based on this history, COMBIgor brings the experience of almost 20 years of software development at NREL for handling and processing combinatorial data. Additionally, it has been integrated with public data repositories to enable cutting-edge data transparency and openness\unskip~\cite{374021:8337475}.

\subsection{COMBIgor Terminology }
The terminology utilized within the COMBIgor package is meant to reflect the common descriptors used in HTE literature. 
A list of the most important terms is shown in Table \ref{table_terms}. 
For example, it is important to understand the concept of a library in the COMBIgor package. 
A library is a specimen with spatially-varying processing conditions, such as substrate temperature or plasma flux.
A collection of libraries constitutes a project.
Libraries are measured on a mapping grid - a discretized, spatial arrangement of points from which data are collected on a set of instruments.
A single point of the mapping grid, which corresponds to a specific location on a library, is called a sample.
This language is used extensively in COMBIgor for referencing HTE specimens and their corresponding data.

\begin{table*}[!htbp]
\caption{{COMBIgor terminology common to combinatorial literature} }
\label{table_terms}
\def\arraystretch{1}
\ignorespaces 
\centering 
\begin{tabulary}{\linewidth}{p{\dimexpr.2\linewidth-2\tabcolsep}p{\dimexpr.8\linewidth-2\tabcolsep}}
\hline 
Term & Definition\\
\hline 
Library &
  A specimen processed with spatially varying conditions, such as synthesis temperature or compositional flux.\\
Mapping Grid &
  A spatial arrangement of points on a combinatorial library from which data are collected by some set of instruments.\\
Sample &
  A location on a library that corresponds to any single point of the mapping grid. \\
Project &
  A set of combinatorial libraries with a common mapping grid.\\
  \hline 
Data Type &
  A descriptor for a processing variable, a type of data from a measurement, or an experimental observation that characterizes a sample. \\
Meta Data &
  A string variable that describes a sample. Example: process gas used during synthesis. \\
Library Data &
  A numeric descriptor for an entire library. Example: duration of synthesis time.\\
Scalar Data &
  A numeric descriptor for a sample. Example: film thickness.\\
Vector Data &
  A 1D data array for a sample. Example: x-ray diffraction data.\\
Matrix Data &
  A 2D data array for a sample. Example: scanning electron microscope image.\\
\hline 
Add-on &
  An extension to core COMBIgor functionality, such as an instrument or a plugin.\\
Instrument &
  A source of combinatorial data Example: processing or characterization tool.\\
  Plugin &
  A set of operations that process data within COMBIgor.\\
\hline 
\end{tabulary}\par 
\end{table*}

One of the most important conceptual advancements of COMBIgor is that it categorizes the data based on  dimension, scope, and type, regardless of its source, allowing for a greater level of abstraction and code reuse.
Data scope can include a single sample or all samples on a library; data can be numeric or textual; and the numeric data dimension can be scalar, vector, or matrix. 
For example, x-ray diffraction patterns and optical absorption spectra are both numeric vector data types comprising one-dimensional data collected for each sample on the library.
Other important features within COMBIgor are add-ons, which are categorized as either instruments or plugins.
An instrument is a source of synthesis or characterization data from combinatorial experiments, such as XRD or XRF. 
A plugin is a specific set of operations that process data within COMBIgor. 
Examples include background removal or applying mathematical operators.

\subsection{Igor Pro environment }

COMBIgor was written in the Igor Pro environment. Igor Pro\unskip~\cite{374021:8324051} is a commercial software developed by WaveMetrics, Inc. (https://www.wavemetrics.com/), which is well-suited to and commonly used for accessing, processing, storing, and displaying scientific and engineering data in many forms. 
Within the Igor Pro environment, data are stored in array structures called "waves", which can be one to four dimensions comprised of various types of data (text, integers, doubles, bytes, etc.).
To promote organization, waves are stored within nested folders that can be explored and rearranged through the Igor Pro data browser.
Folders and their contents are saved as an Igor Pro experiment, which also houses any other user-created content, such as figures, tables, or notebooks.
The native Igor Pro environment contains numerous built-in functions for plotting, fitting, and performing statistical analysis, in addition to a plethora of extensions and packages for additional, more specialized functionality. 
In addition to Igor Pro's many built-in features, its programming language  enables the development of custom, automated, and complex functionality. 
While Igor Pro is not the only option with such flexible capabilities, it was chosen as the best option for COMBIgor development because it is easily accessible, has features essential for HTE data, includes extensive documentation, and has a legacy of use at NREL. 

\bgroup
\fixFloatSize{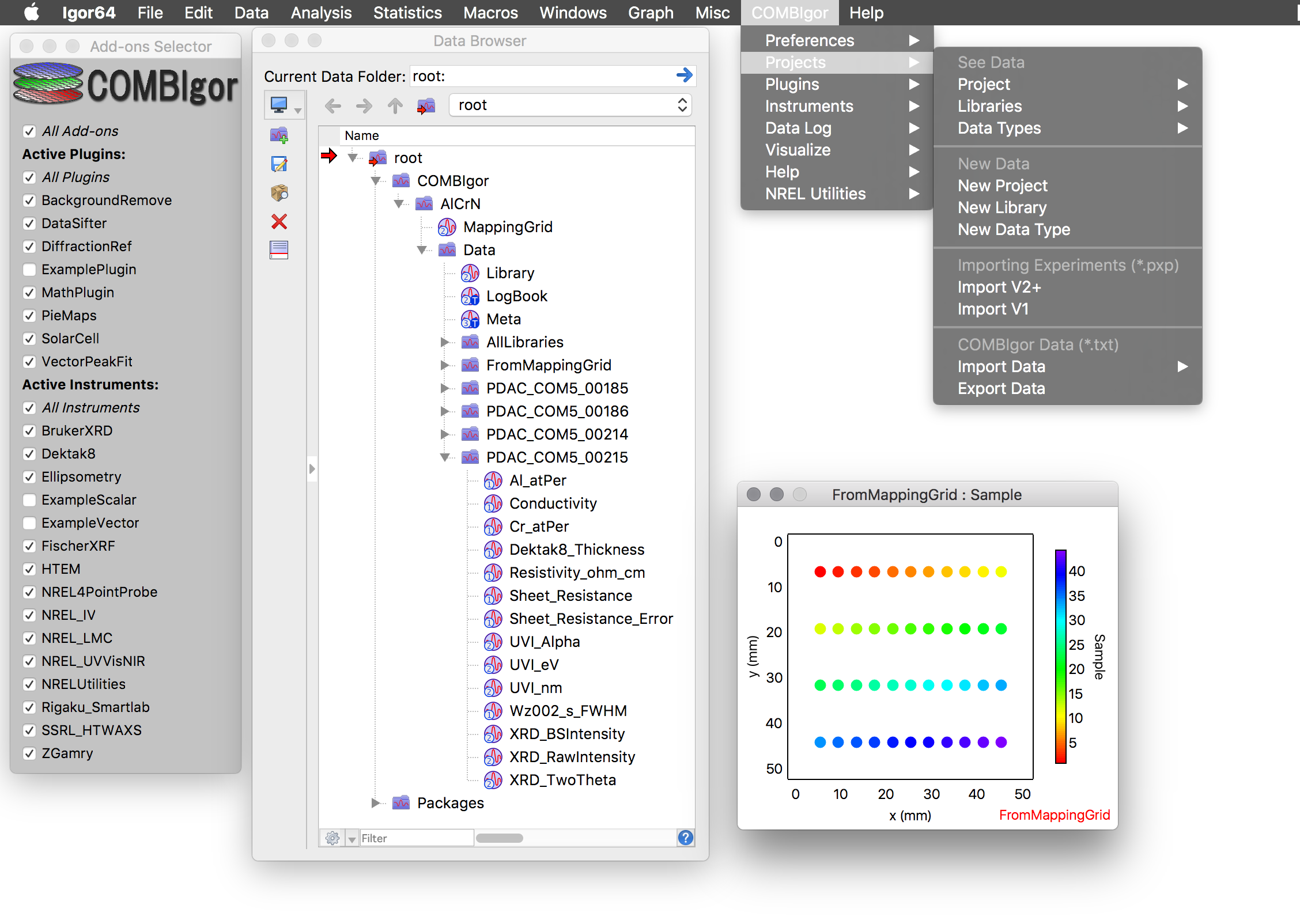}
\begin{figure*}[b!]
\centering \makeatletter\IfFileExists{images/MainScreen.png}{\includegraphics{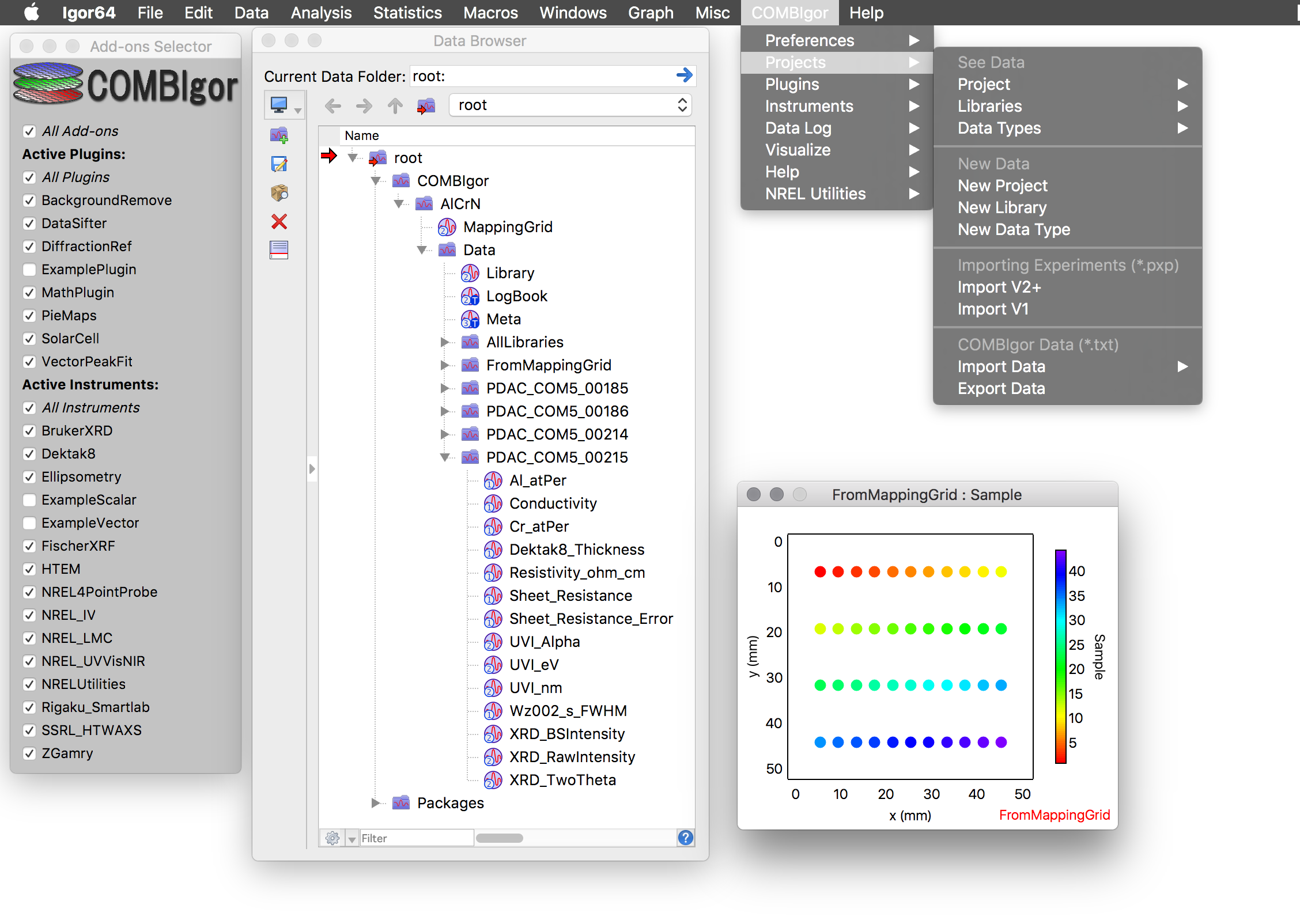}}{}
\makeatother 
\caption{{Common elements of the COMBIgor package (left to right): the  add-on selector for Plugins and Instruments, data browser for the project used in this demonstration, plot of the standard NREL library mapping grid defined for the project, and the COMBIgor user menu showing the project level data operations.}}
\label{figure_basicCombigorStructure}
\end{figure*}
\egroup

\subsection{Distribution and Use}

At time of publication, the COMBIgor package is housed and maintained through a public repository on GitHub \BreakURLText{(https://github.com/NREL/COMBIgor/)} under an open-source license. 
The latest version can be downloaded there and updates can be added through a pull request.
More information about the most recent updates, use cases, download process, bug reporting, etc. can be obtained at \BreakURLText{https://www.COMBIgor.com}. 
The description below is specific to COMBIgor version 2.4. Later versions are likely to have more add-ons as COMBIgor expands after publication of this manuscript. 

After installation (described in section 3.1), the COMBIgor package exists within the Igor Pro environment and can be activated at any time through the native Igor Pro "Data" menu. 
Once activated by the user, the COMBIgor procedure files are mounted into the experiment, the user menu is created, and the add-ons selector is displayed.
After the desired add-ons are selected and the project (with its mapping grid) is defined, the consequent workflow varies depending on the needs of the user. 
Generally, it involves repeatedly loading, processing, plotting, and exporting various types of HTE data. 
A snapshot of COMBIgor package elements within the Igor Pro environment can be seen in Figure~\ref{figure_basicCombigorStructure}. 
More details on the COMBIgor workflow are provided below.

\subsection{Workflow}

The generalized workflow for using COMBIgor is presented in Figure~\ref{figure_workFlow}. 
After COMBIgor package activation, the first six steps ("Setting Up COMBIgor" in Figure~\ref{figure_workFlow}) define the project space, activate the desired add-ons, and initialize the parameters necessary to use them.
A project is defined by a common combinatorial mapping grid for all libraries contained within that project. 
The user must then define loading parameters related to the individual instruments, such as data type names, location of data in the files, and processing-related variables. 
Unlike the mapping grid, these can be redefined during the course of the project lifetime, although these steps are only required when starting a new project.  
Subsequently, data are managed through a project, library, and data type hierarchy.

\bgroup
%\fixFloatSize{images/ProcessFlow.png}
\begin{figure*}[h]%[!htbp]
\includegraphics[width=0.5\textwidth]{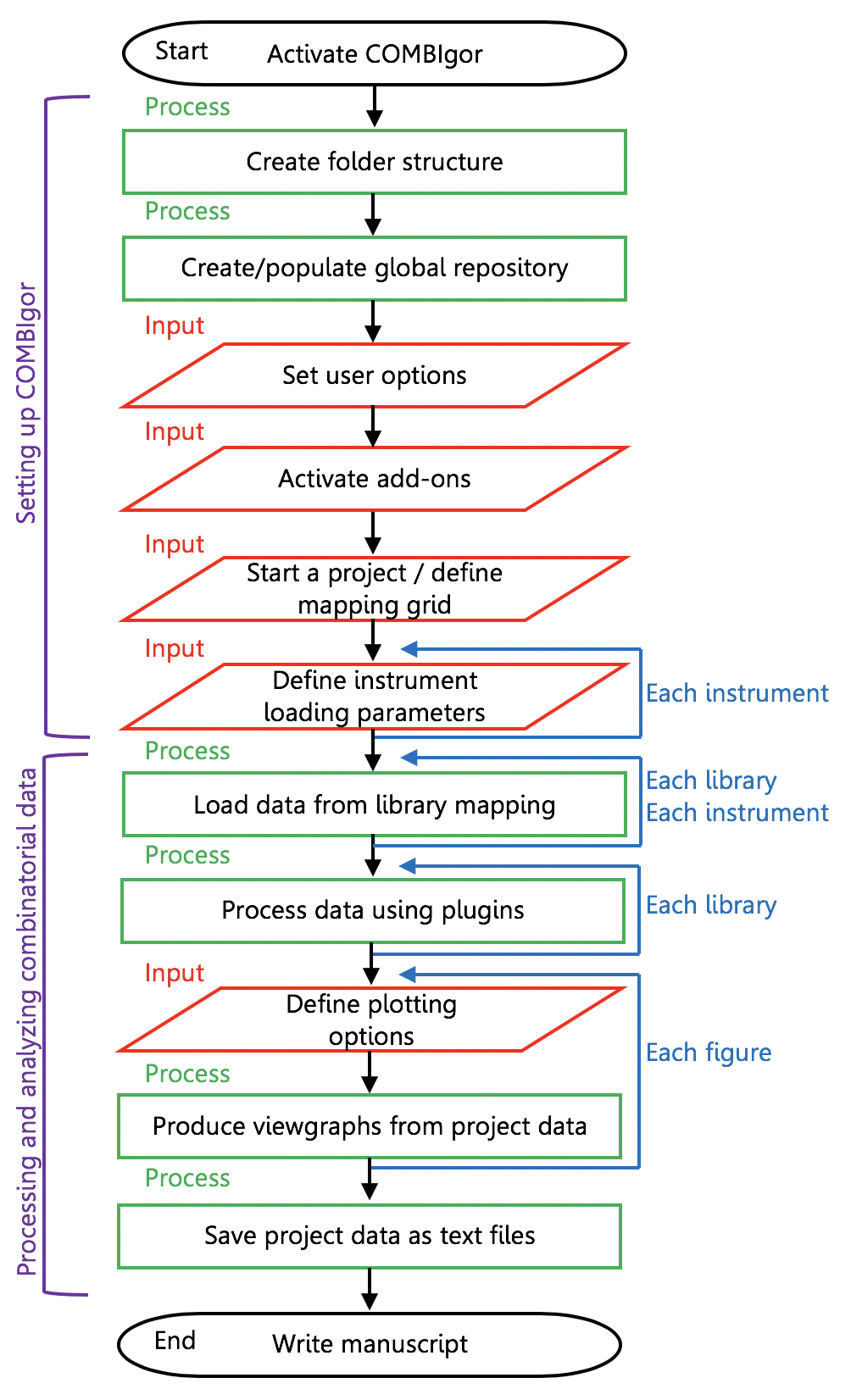}
\makeatother 
\caption{{Typical COMBIgor work flow for loading, processing, and displaying data from HTE libraries. The first six steps, after activation, initialize the project work space. 
The last five steps are repeated, as needed, by the user to analyze all experimental results.
The project can be saved and reopened at any point in the flow. 
Data can be exported into delimited text files and imported into future COMBIgor projects for maximum flexibility of the work flow.}}
\label{figure_workFlow}
\end{figure*}
\egroup

The last five steps ("Analysis of combinatorial data",  Figure~\ref{figure_workFlow}) are executed repeatedly as the scientist loads and analyzes experimental data. 
At this stage, the user loads combinatorial data collected from instruments using the standard mapping grid into the user-defined library folders. 
Once data are loaded into the COMBIgor project, they can be analyzed and visualized using instrument-specific routines, plugins, or native Igor Pro routines.
The number of libraries loaded, instruments utilized, and processing steps will vary depending on the project and the user, as reflected in the loops of Figure~\ref{figure_workFlow}. 
During this process, COMBIgor data loading and processing events are logged into a searchable log book. 
At any point, the raw data can be exported for distribution into tab-delimited text files for loading into additional COMBIgor experiments or other programs, as needed. 
This exporting capability is particularly important for multi-institutional projects, such as the High-Throughput Experimental Materials Collaboratory\unskip~\cite{374021:8324933}. 
All figures generated in COMBIgor can be further modified by the user and exported for use in presentations and manuscripts.

\section{Functionality}
To demonstrate the functionality of COMBIgor, data from combinatorial Al$_{1-x}$Cr$_{x}$N thin film sample libraries are used in the following sections. 
This material system has been recently investigated for structural evolution, electrical, and optical properties \unskip~\cite{374021:8586428}. 
The Al\ensuremath{_{1-x}}Cr\ensuremath{_{x}}N heterostructural alloys were deposited from elemental Al and Cr targets in the presence of nitrogen via radio frequency co-sputtering.
The compositions of these combinatorial libraries were determined by x-ray fluorescence (XRF), structures by x-ray diffraction (XRD), thicknesses by step-edge profilometry, optical properties by optical transmission/reflection spectroscopy, and electrical conductivity by 4-point probe sheet resistance measurements.
More details regarding the deposition or characterization instruments can be found elsewhere\unskip~\cite{374021:8586454}.

\subsection{Installation and Preferences}

The installation process for the COMBIgor package is designed to be as simple as possible. 
An installer experiment file (COMBIgor\_installer.pxp) is included in the package folder. When opened, this file will install the package contents and any necessary Igor Pro extensions. 
The user can install the package with a single click from the COMBIgor menu. 
The installer will preserve a copy of any COMBIgor package files that predate the current installation to protect any modifications a user has made to a previous package version. 
After installation, the user is directed to COMBIgor activation in the data menu, 
%are they actually directed to all of these things?
as well as the COMBIgor help menu and the included tutorial.
In addition, we recommend that before using COMBIgor, the users familiarize themselves with the built-in Igor Pro functionality by taking the guided tour from the Igor Pro help menu. 

The COMBIgor package has a built-in user preference system. 
The user can select a set of COMBIgor preferences, including setting a default import or export folder location, controlling the "kill" behavior of windows generated by Igor (e.g. figures, tables), and generating example programmatic call lines for operations performed by COMBIgor in the command history. 
The package can also generate standard plots upon loading data from instruments and automatically adjust font styling of control panels and plotted figures.

\subsection{Instruments and plugins}
The needs of  HTE vary greatly between projects, users, and laboratories. 
If a single package contained all possible data loading, processing, and visualization functionality applicable to HTE, the package would be too complex to be efficiently navigated.  
For this reason, the COMBIgor package allows each user to select only the optional capabilities (add-ons) that they plan to use.
In addition to keeping the package streamlined,  instrument and plugin add-ons provide a mechanism for individual user expansion and laboratory-wide instrument collections. 

The lists of the base instruments and plugins included at the time of publication of this manuscript are described in Table~\ref{table_instruments} and  Table~\ref{table_plugins}, respectively.
While the plugins are generally useful regardless of the data source, the instruments included in COMBIgor are specific to NREL. 
However, some of these instruments are used at other laboratories that employ HTE methods.\cite{Hattrick-Simpers2019,Ruiz-Yi2016,Green2017a,Miller2014}
Also included in the COMBIgor package are well-commented, generic examples of code needed to create a plugin, a scalar-data instrument, and a vector-data instrument, provided as a starting point for new add-on development.

\begin{table*}[!htbp]
\caption{{Instruments included in COMBIgor} }
\label{table_instruments}
\def\arraystretch{1}
\ignorespaces 
\centering 
\begin{tabulary}{\linewidth}{p{\dimexpr.3131\linewidth-2\tabcolsep}p{\dimexpr.6869\linewidth-2\tabcolsep}}
\hline 
Instrument Name & Description of Data Files\\
\hline 
ExampleScalar &
  Generic example instrument for loading a 2D file of scalar data from library mapping. \\
ExampleVector &
  Generic example instrument for loading a folder of vector data files from library mapping. \\
  \hline 
BrukerXRD &
  X-ray diffraction patterns from a Bruker D8 Discover diffractometer (Diffractplus .raw format).\\
Rigaku\_Smartlab &
  Output files from a Rigaku x-ray diffractometer for XRD and XRR measurements (.ras format).\\
Dektak8 &
  Profilometer scans across step-edge features on a library collected on a Veeco Dektak8 profilometer (.csv format).\\
FischerXRF &
  Composition and thickness from a Fischer x-ray fluorescence instrument (.exp format).\\
ZGamry &
  Output electrochemical impedance analysis files from Framework software by Gamry.\\
Ellipsometry &
  Absorption as a function of wavelength data output from the CompleteEase software (.txt format) \\
    \hline 
NREL4PointProbe &
  Sheet resistance data from a custom-made 4-point probe mapping system at NREL (.hdf5 format).\\
NREL\_IV &
  Folders of current vs. voltage measurements from a custom-built electrical probing station at NREL (.csv format).\\
NREL\_UVVisNIR &
  Optical characterization from a custom-built thin film transmission and reflection system at NREL (.hdf5 format).\\
SSRL\_HTWAXS &
  Corrected detector matrix data (.mat format) or column-averaged vector data (.csv format) from the high-throughput wide-angle x-ray scattering beamline (1-5) at the Stanford Synchrotron Radiation Lightsource at SLAC. \\
 \hline 
 HTEM &
  Data retrieved from the High Throughput Experimental Materials (HTEM) database.\unskip~\cite{374021:8337475}\\
  NREL\_LMC &
  Output files from the Laboratory Metadata Collector (LMC) used at NREL for deposition conditions (.json format).\\
\hline 
\end{tabulary}\par 
\end{table*}

\begin{table*}[!htbp]
\caption{{Plugins included in COMBIgor} }
\label{table_plugins}
\def\arraystretch{1}
\ignorespaces 
\centering 
\begin{tabulary}{\linewidth}{p{\dimexpr.3062\linewidth-2\tabcolsep}p{\dimexpr.6938\linewidth-2\tabcolsep}}
\hline 
Plugin Name & Description of functionality\\
\hline 
ExamplePlugin &
  An example of a COMBIgor plugin designed as a starting point for development.\\
\hline 
BackgroundRemove &
  Masks, fits, and subtracts backgrounds from vector data, such as x-ray diffraction.\\
DiffractionRef &
  Imports, stores, and plots powder diffraction references exported from common diffraction programs and databases.\\
VectorPeakFit &
  Executes constrained peak fits on a library of vector data and extracts fit information.\\
DataSifter Plugin &
  Applies a conditional test and, upon true, performs some resulting action across every sample on a library.\\
MathPlugin &
  Performs simple mathematical operations to data across every sample on a library.\\
Piemaps &
  Plots library maps on which each sample marker is a pie chart of multiple scalar data types.\\
TernaryPlotter &
  Generates ternary plots with marker color and size determined by scalar or vector data.\\
SolarCell &
  Extracts solar cell performance parameters from current vs. voltage data.\\
\hline 
\end{tabulary}\par 
\end{table*}

\subsection{Data Loading and Processing}
Experimental data comes in many different file formats: some are stored as a simple text file, some are contained in proprietary file formats, and most are something in between. 
Taking advantage of the data-handling features of Igor Pro, COMBIgor is able to import a wide variety of data formats.
Data loading involves programmatically opening native instrument file types, transferring the data for each sample into an Igor Pro wave that matches the dimensionality of the incoming data, and storing those waves in the appropriate project and library folders.
Loading steps include assigning identifying information, such as project, library, sample index, data type, and data dimension, to the incoming data; creating the data storage in the correct location within the COMBIgor experiment; and populating that storage with the appropriate data. 
The load event is documented in the project log book to retain a record of the original data source.

\bgroup
\fixFloatSize{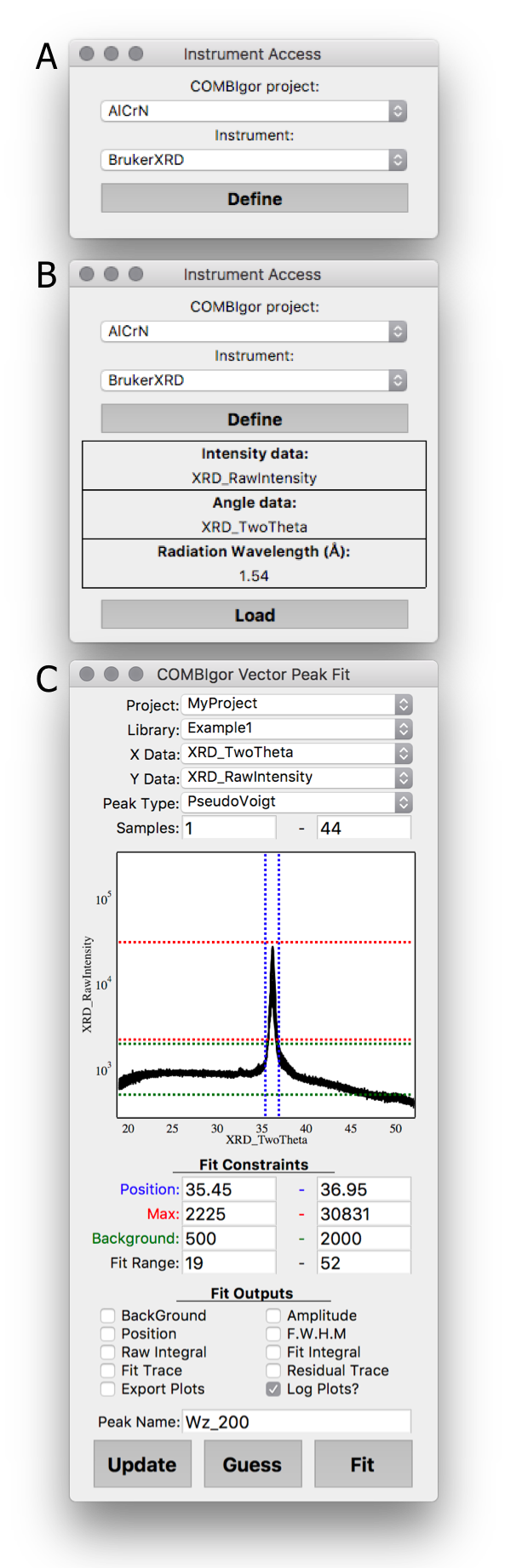}
\begin{figure*}[!htbp]
\centering \makeatletter\IfFileExists{images/3Panel.png}{\includegraphics[width=.35\linewidth]{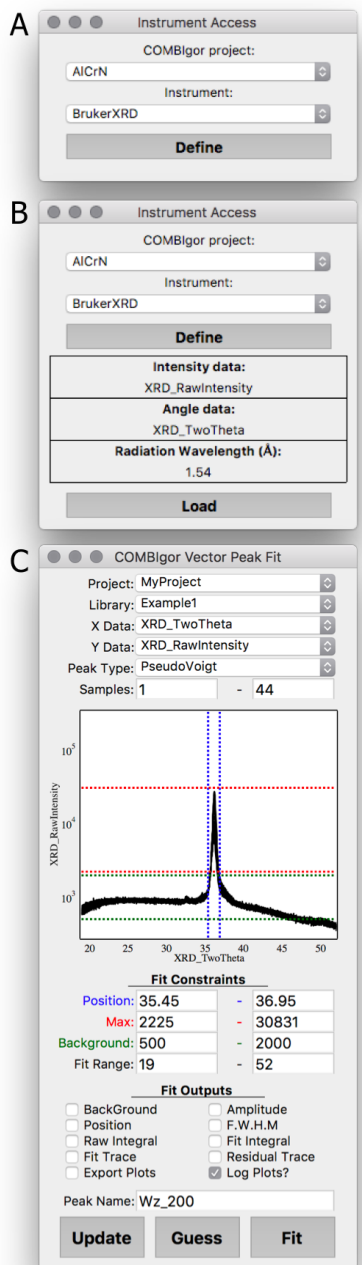}}
\makeatother 
\caption{{Instrument access panel used to define data types and variables necessary to load data from the selected instrument into the selected project, shown (A) prior to instrument definition, where the option to load is not given, and (B) after instrument definition, where instrument globals have been defined and the load option is shown. (C) COMBIgor vector peak fitting plugin demonstrating peak fitting and integration of the wurtzite (002) x-ray diffraction peak for all 44 samples of a single Al\ensuremath{_{1-x}}Cr\ensuremath{_{x}}N combinatorial library.}}
\label{figure_accessPanel}
\end{figure*}
\egroup

In COMBIgor, data are typically loaded through instrument add-ons (Figure~\ref{figure_accessPanel}).
Additionally, data can be imported from other COMBIgor files as COMBIgor text files. 
Finally, COMBIgor has the ability to fetch data directly from online databases through application programming interfaces (APIs) and store it locally for comparison to other data types. 
This feature is demonstrated by loading data from the High Throughput Experimental Materials (HTEM) database (https://htem.nrel.gov),\unskip~\cite{374021:8337475} which houses experimental data generated by the HTE mapping instruments at NREL. 
With the use of the HTEM instrument and an internet connection, an HTE researcher can import data from any tool in only a few clicks.
Data processing is integral to scientific experimentation. 
For the HTE researcher, the diversity of processing tasks combined with the magnitude of collected data can create a daunting workload. 
For this reason, COMBIgor data is extremely easy to batch-process, and can be processed in various modes. 
There are four ways in which COMBIgor data can be processed:

  \begin{itemize}
  \item \relax \textbf{Upon loading by an individual instrument procedure}\\
  This is the simplest method, and is utilized for standard measurement processing routines. \textit{Example: Feature height extraction from profilometry scans of thin film samples.}
  \item \relax \textbf{By COMBIgor plugins}\\
  This is the most powerful method for HTE with high density mapping grids, since the same process can be systematically applied to every sample on a library. \textit{Example: Peak fitting to extract the FWHM of an x-ray diffraction peak. See Figure \ref{figure_accessPanel} for more detail.}
  \item \relax \textbf{Utilizing native Igor Pro functionality}\\
  This method is the most flexible, since the built-in capabilities of Igor Pro are very extensive. Anything that can be done in Igor Pro can be developed into a plugin for a specific task. \textit{Example: Using the Igor Pro function `Differentiate' to identify peaks in an x-ray fluorescence spectrum.}
  \item \relax \textbf{Additional Igor Pro packages}\\
  Numerous user-developed data processing packages exist for Igor Pro which, if loaded alongside COMBIgor, can be used to process COMBIgor data. \textit{Example: Fitting x-ray reflectivity patterns using the Motofit\unskip~\cite{374021:8356201} Igor Pro package.}
  \end{itemize}

\subsection{Data Visualization} 
Once data are loaded, the COMBIgor package can produce publication-quality graphics. 
One of the strengths of COMBIgor is its ability to plot data types of different dimensionality on the same plot. 
There are four modes of graphics generation:

\bgroup
\fixFloatSize{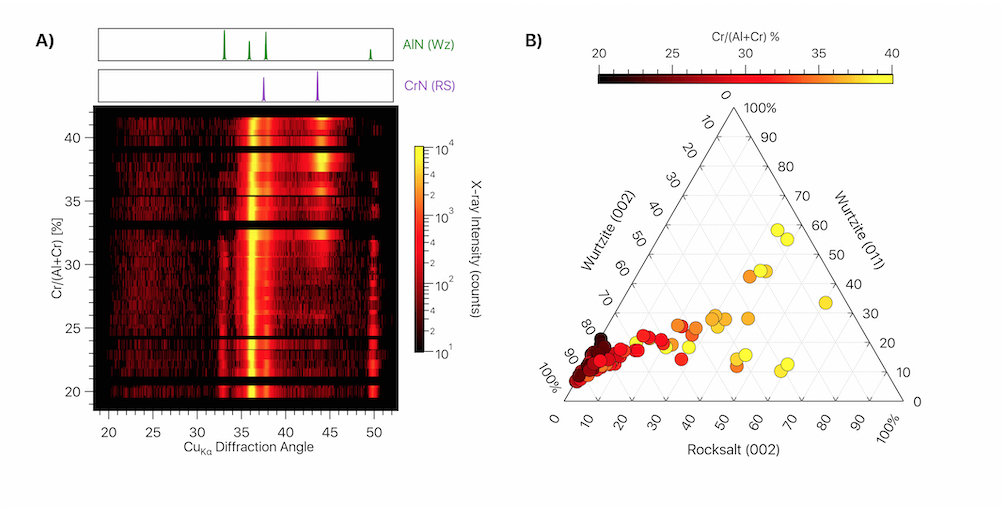}
\begin{figure*}[!htbp]
\centering \makeatletter\IfFileExists{images/XRD_Ternary.jpg}{\includegraphics{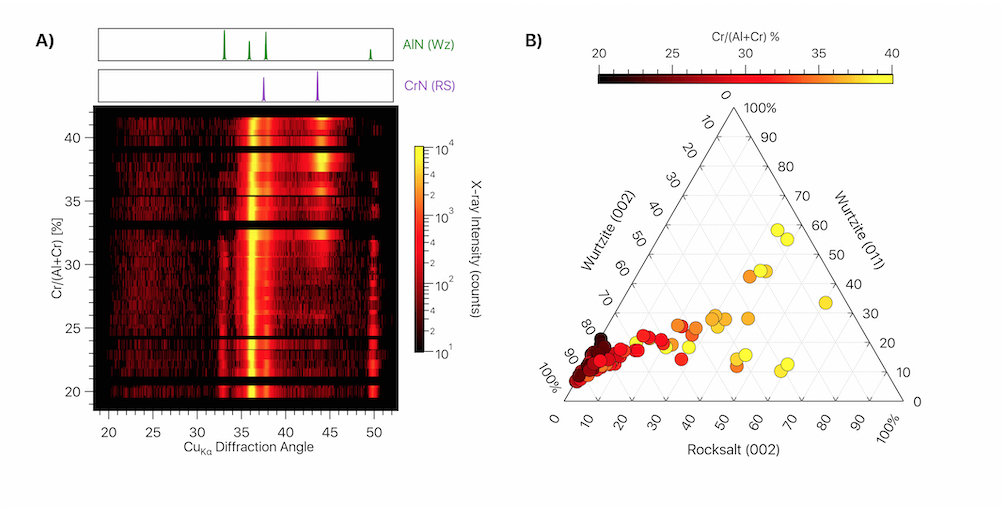}}{}
\makeatother 
\caption{{COMBIgor graphics generated from a single Al\ensuremath{_{1-x}}Cr\ensuremath{_{x}}N combinatorial library. 
(A) Heat map of x-ray diffraction data (color axis) as a function of composition data (vertical axis). 
Data were imported by COMBIgor. The background was subtracted using the BackgroundRemove plugin. 
Data were plotted with the COMBIgor Display panel, and reference profiles of wurtzite-structured AlN and rocksalt-structured CrN were added with the DiffractionRef plugin. 
(B) Ternary plot of x-ray diffraction peak area distributions. 
Peak areas from the profiles shown in part A were extracted using the VectorPeakFitting plugin (Figure~\ref{figure_accessPanel}) and mapped using the TernaryPlotter plugin.}}
\label{figure_combigorGraphics}
\end{figure*}
\egroup

  \begin{itemize}
\item \relax \textbf{Upon instrument-specific loading}\\
If the user selects the option in package preferences, COMBIgor can automatically generate formatted plots as data are loaded through the instrument procedures. \textit{Example: Programmatic generation of a normalized diffraction intensity heat map comparing all samples upon loading X-ray diffraction data (see Figure~\ref{figure_combigorGraphics}, part A).}
\item \relax \textbf{Using the COMBIgor Display panel}\\ 
The COMBIgor display panel is capable of producing two-dimensional figures with up to four data type-dependent axes: horizontal axis, vertical axis, marker color, and marker size. Any data dimensionality can be handled by each of these axes. \textit{Example: The Display panel can be used to generate a map of material resistivity for an entire library, on which the marker size for each sample is determined by film thickness (see Figure~\ref{figure_absorptionData}, part A).} 
\item \relax \textbf{Utilizing native Igor Pro functionality}\\ 
Data organized within a COMBIgor project can be utilized directly by Igor Pro graphics functionality to produce fully customized, publication-quality graphics. \textit{Example: Adding a visible spectrum color indicator to the axis of a COMBIgor generated 2D absorption coefficient plot greatly increases the visual appeal of this graphic (see Figure~\ref{figure_absorptionData}, part B.)}
\item \relax \textbf{Visualization plugins}\\ 
Additional plugins extend COMBIgor functionality to specialized plot types. Plugins can be created and added to COMBIgor to generate any graphic type Igor Pro can produce. \textit{Example: Using the TernaryPlotter plugin to analyze the distribution of peak area as determined by the VectorPeakFitting plugin (see Figure~\ref{figure_combigorGraphics}, part B).} 
  \end{itemize}

\bgroup
\fixFloatSize{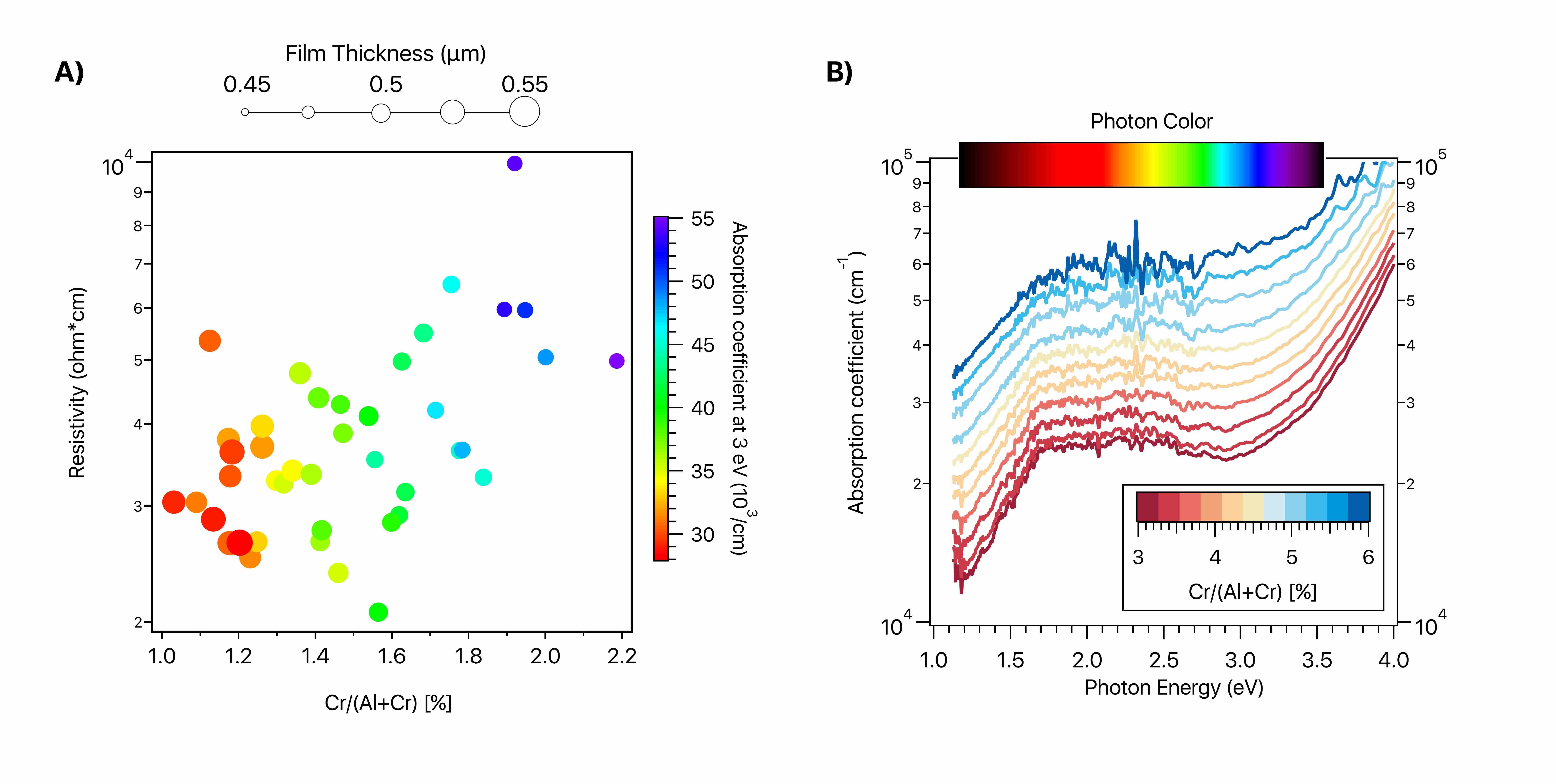}
\begin{figure*}[!htbp]
\centering \makeatletter\IfFileExists{images/Electrical_Optical.jpg}{\includegraphics{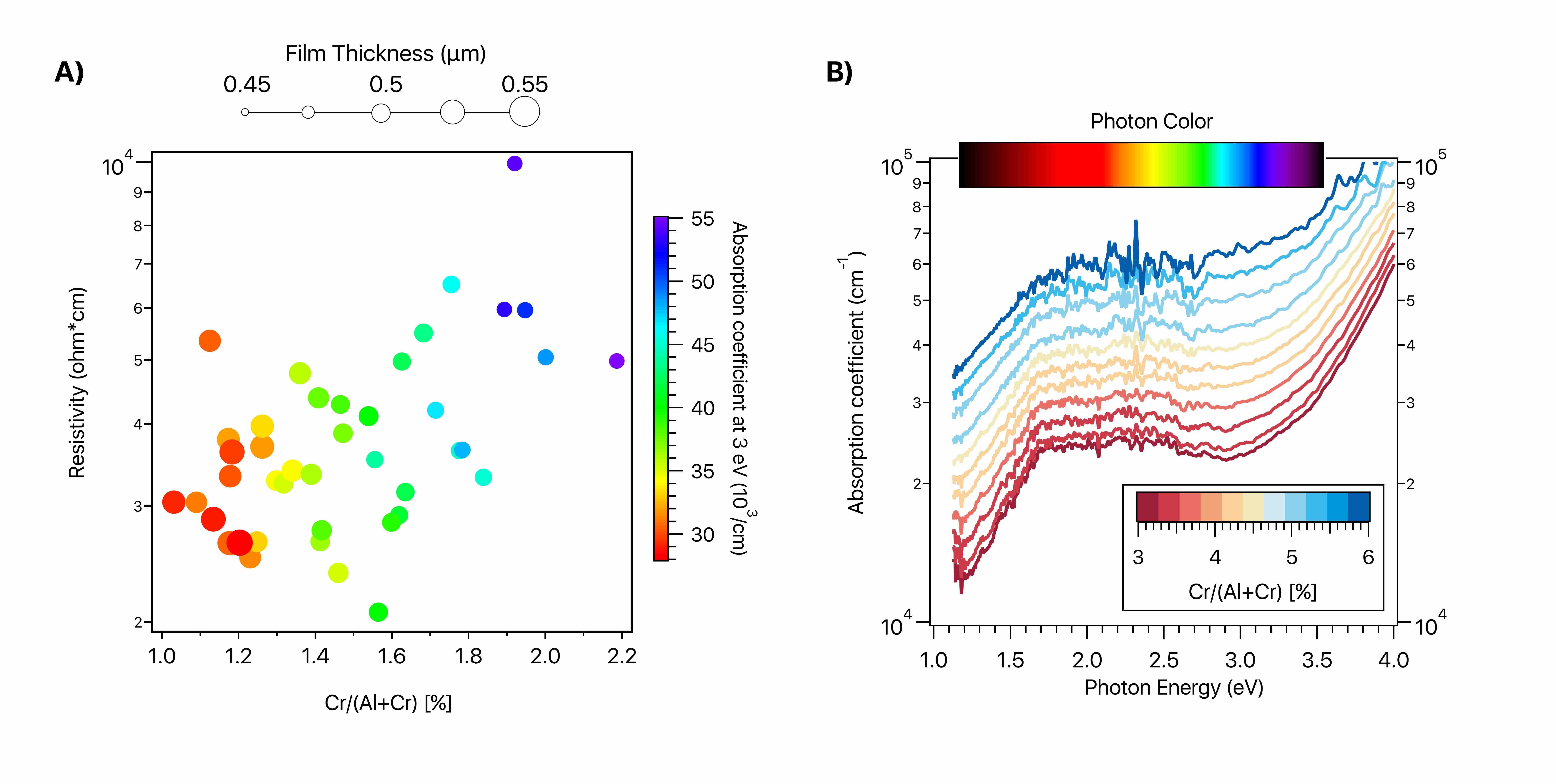}}{}
\makeatother 
\caption{{COMBIgor graphics generated from a single Al\ensuremath{_{1-x}}Cr\ensuremath{_{x}}N combinatorial library.  
(A) Comparison of film resistivity, composition, thickness, and absorption coefficient at 3 eV produced by COMBIgor display core functionality. 
Resistivity was imported and processed by the NREL\_4PointProbe instrument. 
Film thickness values were imported and processed through the Dektak8 instrument. 
Chromium cation percent was imported and sorted by the FischerXRF instrument. 
Absorption coefficient at 3 eV was extracted using the COMBIgor DataSifter plugin.
B) Absorption coefficient as a function of energy with an inset of the visible range of the electromagnetic spectrum, highlighting changes in the absorption characteristics as the Cr cation fraction changes. 
Transmission and reflection data were imported and processed into absorption coefficient using the NREL\_UVVisNIR instrument.
All instruments and plugins are part of the COMBIgor package. }}
\label{figure_absorptionData}
\end{figure*}
\egroup

The magnitude of data generated by HTE can lead to comparison graphics that contain thousands of data points. 
Graphics produced by the display features of COMBIgor are connected to a display wave specifically created for this visualization and distinct from the original data waves.
The source information of the wave data (e.g. library name, data type, data dimension, sample range, etc.) can be accessed and easily saved.
However, this also means that the plot has to be remade when the data are updated or reloaded (i.e. the plot is not updated automatically).
When plots are created, they can be saved to the export folder and recreation macros can be generated. 
Closing the plot will trigger deletion of the attached display wave, thus minimizing data duplicity. 

\subsection{Data Access and Porting}
COMBIgor data processing and storage is designed to be transparent and easy to access so that the user understands the process and can modify the approach at will. 
To this end, all COMBIgor data are stored within the folders of the Igor Pro experiment, as shown in Figure~\ref{figure_basicCombigorStructure}. 
Metadata and library data types are stored in a single wave per project, which facilitates easy library-to-library data comparison. 
Scalar, vector, and matrix data types are stored as individual waves in library-specific folders within the project folder for easy sample-to-sample comparisons. 
Any of these waves can be easily accessed, giving the user the power to inspect the numbers and text directly. 
The project user menu (Figure~\ref{figure_basicCombigorStructure}) allows the user to inspect the projects, libraries, and data types stored in the experiment by COMBIgor. 
When selected from the project menu, the data are presented in the Igor Pro data browser.

Combinatorial HTE is often collaborative. Results 
are published in peer-reviewed journals that encourage submission of raw data used in the publication, either as supporting information or in a separate data repository.
Journals that publish mainly data, such as Data in Brief and Scientific Data, are also becoming more common.\cite{DataInBreif,Morebangforyourbyte}
For all of these reasons, data stored in COMBIgor must be accessible beyond Igor Pro in a simple file format. 
COMBIgor can automatically save any locally-stored data onto the hard drive in a simple tab-delimited text file format with a folder structure that matches the project. 
As mentioned above, these exported data can also be imported into a different COMBIgor project, while maintaining the library names and sample numbers.

\subsection{Documentation}
A major strength of the COMBIgor package is the breadth and depth of documentation, which is designed to be easily accessed at all stages of use and development. 
The documentation is compiled using the Igor Pro help file type, as well as in-line procedure comments. The documentation exists at seven levels:

  \begin{itemize}
  \item \relax A beginner's tutorial is included in the help menu and presented upon installation for new COMBIgor users to learn the basics of instruments, plugins, and display features. The package includes example data for each of the instruments, which are used over the course of the tutorial. While example data are specific to NREL, this is intended to build a conceptual understanding of the work flow.
  \item \relax General information about the mapping grid, data type dimensions, plugin operations, and instrument details are included in a general help file which gives useful details about COMBIgor and can be accessed through the COMBIgor menu.
  \item \relax Developer's documentation exists for all functions not specific to instruments or plugins (except examples), including a description of the function's purpose, as well as inputs and outputs. Programming documentation is provided to enable maximum expandability of the COMBIgor package by users and research groups outside of NREL. 
  \item \relax In-line code documentation, as per standard coding practice, exists within the procedure files. These are for anyone trying to troubleshoot, understand, or expand upon the existing COMBIgor package functions.
  \item \relax An introductory PDF that explains terminology, installation, and activation of COMBIgor is provided in the package folder.
  \item \relax A series of tutorial videos are hosted on \BreakURLText{http://www.youtube.com} and can be found on the tutorial section of the support website (\BreakURLText{http://www.COMBIgor.com}).
  \item \relax This manuscript
  \end{itemize}
    
\section{Package Architecture}
The COMBIgor package is written in the Igor Pro scripting language, which resembles C++, and exists in a single distribution folder. 
COMBIgor was developed on the Apple Macintosh operating system and has been tested on both Apple Macintosh and Microsoft Windows operating systems. 
The open-source COMBIgor operates within the commercial Igor Pro environment which is operating system neutral.
It is designed to be contained in as few procedure files as possible with each procedure file containing multiple functions.  
The procedure files are grouped in four folders: 
 \begin{enumerate}
  \item \relax The \textit{Procedures} folder contains the procedures needed to perform all core package functionality.
  \item \relax The \textit{Instruments} folder contains a single procedure for each instrument. Each of these procedures contains the unique functions needed to load and process data from that instrument. 
  \item \relax The \textit{Plugins} folder contains a single procedure file for each plugin, which builds its panel interface and performs the data manipulation processes. 
  \item \relax The \textit{Help} folder contains the package documentation for learning, using, and programming in COMBIgor. 
  \end{enumerate}
  
Tasks performed by the package elements can be categorized into three categories: core functionality, data handling, and add-ons. 
For each of these three categories, the package elements that perform or support the sub-tasks are illustrated in Figure~\ref{figure_structureSchematic} and described in more detail below.

\bgroup
\fixFloatSize{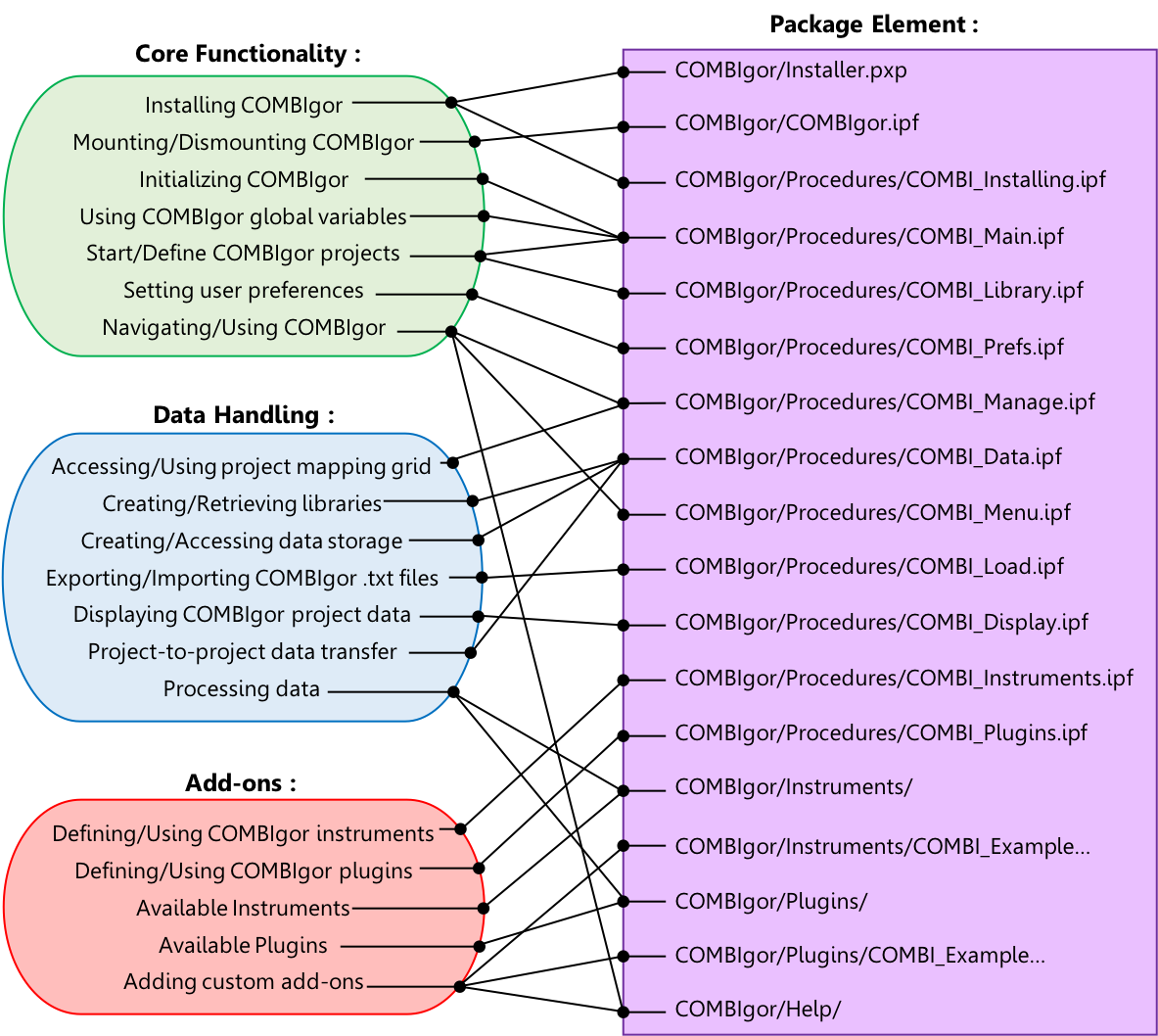}
\begin{figure}[t!]
\centering \makeatletter\IfFileExists{images/WebOfRelations.png}{\includegraphics{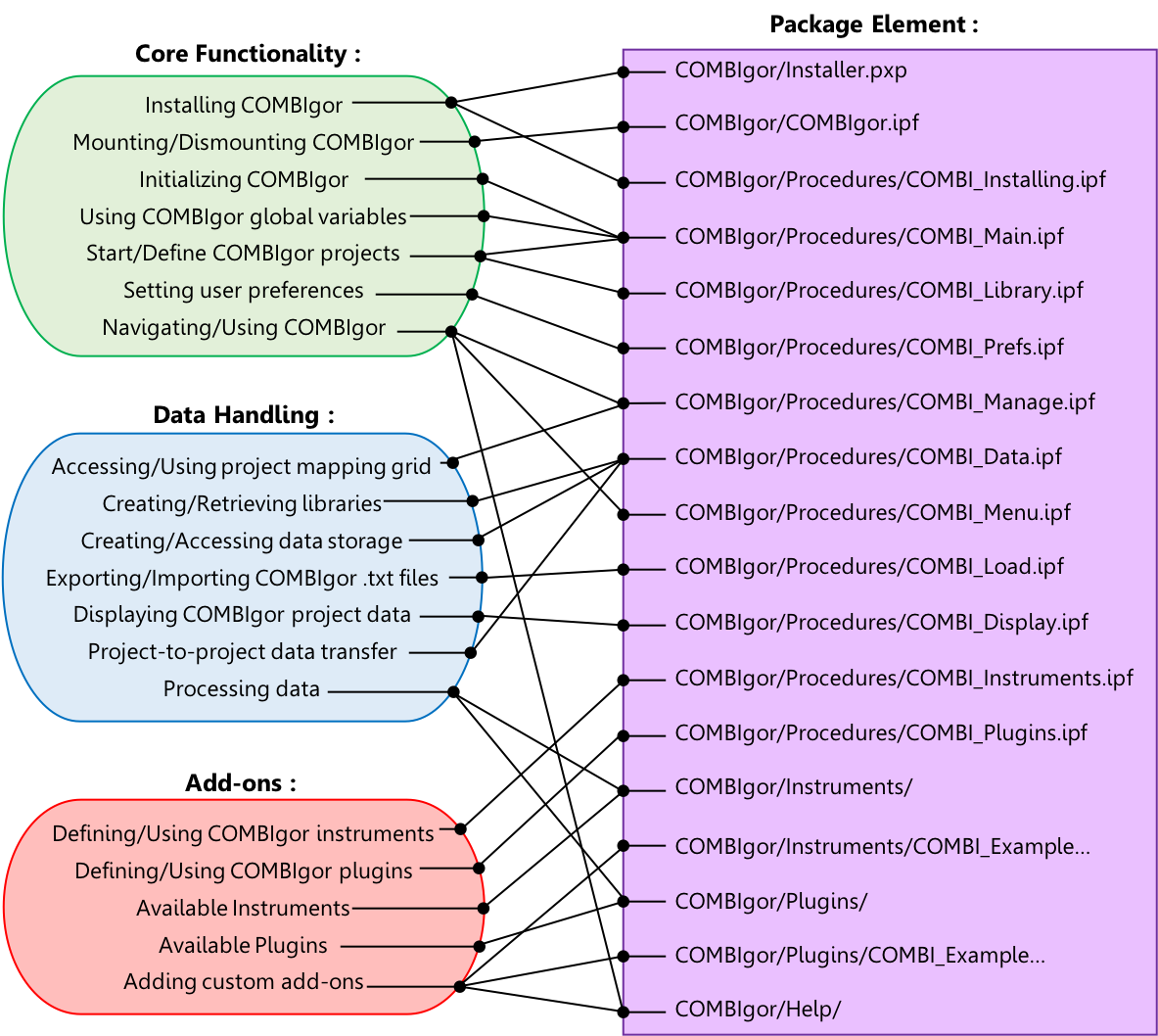}}{}
\makeatother 
\caption{{Schematic representation of the COMBIgor package elements and programmatic tasks contained and executed by each. 
Tasks are categorized into three groups (left) and linked to the package elements (right) which are designed to perform or aid the task.}}
\label{figure_structureSchematic}
\end{figure}
\egroup

\subsection{Core functionality}
COMBIgor core functionality includes installing, mounting, initializing, using, and navigating the package. 
Installation of the package involves placement of two aliases into the Igor Pro user folders and installation of built in XOPs. 
These steps can be done manually or by the COMBIgor\_installer.pxp experiment.
After installation, the user can activate COMBIgor from the data menu, enabling the remaining core functionality.
These functions handle the storage and retrieval of global variables necessary for the package as well as for each individual project. 
Additional global variables can be added by future developers. 
The core functionality builds and enables the Projects menu featured in Figure~\ref{figure_basicCombigorStructure}, which enables the user to navigate the data stored in COMBIgor, create new projects, libraries, and data types, and import and export data from the project. 

\subsection{Data handling}
Data handling tasks within COMBIgor are the most robust, and therefore complex, functions. 
These functions create data storage waves, move data in and out of these waves, and transfer data in and out of the Igor Pro experiment. 
In addition to data management, these functions perform the vital roles of displaying and processing data. 
Because directly inspecting data in an array with more than two dimensions is difficult, data are typically parsed and stored in one- or two- dimensional waves. 
Each data type is a separate wave and is categorized by the project, library, and data dimensionality. 
The storage waves are of the Igor binary and Igor text file types, which are compressed and stored in an Igor packed experiment file. 
This results in an efficient use of memory when compared to the native instrument files provided as example data.
The most essential data used for the example experiment occupy 80\% less memory when stored in an Igor Pro packed experiment and 50\% less memory when stored as COMBIgor text data files.

\subsection{Add-on packages}
Add-on package elements are the most diverse functions due to the wide variety of tasks they perform. 
By default, COMBIgor has the functions needed to initialize and run any add-on included in the Plugins and Instruments folders. 
Each plugin and instrument listed in Table~\ref{table_plugins} and Table~\ref{table_instruments} has documentation for users; however, programmer documentation does not exist for these add-ons, as they are unlikely to be modified by individual users (with the exception of the ExamplePlugin, ExampleVector, and ExampleScalar procedure files). 
The ability to expand the package is integral to the success of the COMBIgor platform and new functionality should be added by creation of additional add-ons. 
By placing a new instrument procedure file in the Instruments folder of the package (or plugin file in the Plugins folder), it becomes available for activation by the Add-ons panel. 
This can be done at any time and is explained in the documentation. 
Extensively commented example procedures are included for a plugin, a scalar instrument, and a vector instrument. 
These are provided as a foundation for further user development. 

\subsection{Object-oriented programming}
COMBIgor was not written with an obvious object-oriented programming (OOP) architecture in mind.
However, its hierarchical data structure can be explained in terms relevant to OOP concepts.
At the highest level, each COMBIgor project, in a single Igor experiment, can be thought of as an object of the project class with instance variables defined by the mapping grid and the project level variables.
Within the project, each sample library is a single object of the library class, which inherits sampling definitions from the mapping grid of the project. 
At the lowest level, individual pieces of mapped data, specific to each library, are objects of one of the four HTE data classes (library, scalar, vector, or matrix) with instance variables defined by the instrument parameters. 
Different instrument add-ons are instances of the instrument class. The same is true for plugins and the plugin class.
    
\section{Summary}
Here we have presented the COMBIgor package for Igor Pro which aids in the management and analysis of combinatorial materials science data sets. 
The COMBIgor package aims to decrease the burden of HTE data management, while simultaneously improving data transparency, usage, and sharing. 
By using the open-source COMBIgor package for commercial Igor Pro, researchers can spend a greater portion of their time understanding the scientific outcomes of their experiments and less time moving, sorting, and managing data. 

The functionality of COMBIgor is demonstrated for a few core instruments and plugins used at the National Renewable Energy Laboratory.
To show the extent of the package's capabilities, we used Al$_{1-x}$Cr$_x$N combinatorial HTE sample libraries as an example.
Overall, this article is meant to facilitate COMBIgor adoption by researchers, groups, and laboratories that utilize spatially-resolved mapping techniques. 

Numerous resources have been developed to install, manage, and expand COMBIgor, and the underlying architecture of the package is described.
Extensive documentation resources are also included, so that a user with any amount of experience with Igor Pro can use and expand the functionality to suit their research needs. 
To give COMBIgor a try, please start by visiting \BreakURLText{https://www.COMBIgor.com}.

\section*{Acknowledgments}
This work was authored by the National Renewable Energy Laboratory, operated by Alliance for Sustainable Energy, LLC, for the U.S. Department of Energy (DOE).
A.Z., J.D.P., and S.R.B. acknowledge funding from the Office of Science, Office of Basic Energy Sciences, as part of the Energy Frontier Research Center “Center for Next Generation of Materials Design: Incorporating Metastability” under Contract No. DE-AC36-08GO28308.
K.R.T., V.J., and G.L.B. acknowledge support by the United States National Science Foundation award DMREF-1534503. 
K.H. and I.K. were supported by DOE Office of Energy Efficiency and Renewable Energy (EERE), Solar Energy Technology Office (SETO) and Fuel Cell Technology Office (FCTO).
M.C.P. was supported by the Army Research Office under grant number W911NF-17-1-0051.
A.M. was supported by a CoorsTek Fellowship through the Colorado School of Mines Foundation.
D.M.R. was supported by the Department of Mechanical Engineering at the University of Colorado at Boulder.
None of the opinions contained within are necessarily those of the U.S. Government, Colorado School of Mines, or the University of Colorado. 

\bibliography{article}

\end{document}